\documentclass[11pt]{article}

\usepackage{amsfonts}
\usepackage{amsmath}
\usepackage{amssymb}
\usepackage{amsthm}
\usepackage[mathscr]{eucal}
\usepackage{bbold}
\usepackage{braket}
\usepackage{color}
\usepackage{dsfont}
\usepackage{framed}
\usepackage{jheppub}
\usepackage{mathtools}
\usepackage{physics}
\usepackage[normalem]{ulem}
\usepackage{tensor}
\usepackage{thmtools}
\usepackage{thm-restate}
\usepackage{ulem}
\usepackage{tikz}
\usepackage{soul}
\newcommand{\bb}[1]{\mathbb{#1}}
\def\sn{\mathrm{sn}}
\usetikzlibrary{math} %needed tikz library for setting variables
\newsavebox{\boxr}
\newlength{\boxrw}
\definecolor{gcolor}{RGB}{0,100,0}

\newcommand{\pa}[1]{\left(#1 \right)}

\definecolor{rcolor}{RGB}{200,0,0}

\definecolor{bcolor}{RGB}{10,10,255}

\newcommand{\ignore}[1]{}

\newcommand{\ex}[1]{\mathrm{e}^{#1}}

\renewcommand{\(}{\left(}

\makeatletter\def\@fpheader{~}\makeatother
\usepackage[export]{adjustbox}
\title{On Quantum Information Before the Page Time}
 
\author[1,2]{Jonah Kudler-Flam}
\emailAdd{jkudlerflam@ias.edu}

\author[3,4]{and Yuya Kusuki}
\emailAdd{ykusuki@catech.edu}

\affiliation[1]{School of Natural Sciences, Institute for Advanced Study, Princeton, NJ 08540 USA}
\affiliation[2]{Princeton Center for Theoretical Science, Princeton University, Princeton, NJ~08540, USA}
\affiliation[3]{Walter Burke Institute for Theoretical Physics, California Institute of Technology, Pasadena, CA 91125, USA}
\affiliation[4]{RIKEN Interdisciplinary Theoretical and Mathematical Sciences, Wako, Saitama 351-0198, Japan}

\abstract{While recent progress in the black hole information problem has shown that the entropy of Hawking radiation follows a unitary Page curve, the quantum state of Hawking radiation prior the Page time is still treated as purely thermal, containing no information about the microstructure of the black hole. We demonstrate that there is significant quantum information regarding the quantum state of the black hole in the Hawking radiation prior to the Page time. By computing of the quantum fidelity in a 2D boundary conformal field theory (BCFT) model of black hole evaporation, we demonstrate that an observer outside of an evaporating black hole may distinguish different black holes via measurements of the Hawking radiation at \textit{any} time during the evaporation process, albeit with an exponentially large number of measurements. Furthermore, our results are universal, applicable to general BCFTs including those with large central charge and rational BCFTs. The techniques we develop for computing the fidelity are more generally applicable to excited states in CFT. As such, we are able to characterize more general aspects of thermalization in 2D conformal field theory.
}

\begin{document}
 
% Title Page
 \begin{flushright}
CALT-TH 2022-040
\\
RIKEN-iTHEMS-Report-22
\end{flushright}

\maketitle

\section{Introduction}\label{sec:intro}

The black hole information problem was first characterized by Stephen Hawking as a ``breakdown of predictability in gravitational collapse'' \cite{hawking1975particle, hawking1976breakdown}. The essence of the problem is that the information describing the matter that travels beyond the black hole horizon is lost forever; the black hole evaporates, radiating a spectrum of particles that only depends on the macroscopic thermodynamic parameters, such as mass, charge, and angular momentum, with no dependence on the microscopic details of the quantum state of the collapsing matter. Strictly speaking, this is not a breakdown in predictability in the forward time direction because knowing the state of the universe prior to gravitational collapse deterministically fixes, in semi-classical gravity, the quantum state of the universe after evaporation. However, in the backwards time direction one is unable to predict the early-time quantum state of the collapsing matter even with complete knowledge of the final state due to the entropy-generating, many-to-one map of the forward time evolution. The quantum states of the radiation emitted from distinct black holes are entirely \textit{indistinguishable}. 

The original computations of the quantum state of the radiation were done on fixed spacetime geometries with only minor adjustments included to account for the gravitational backreaction of the radiation \cite{hawking1975particle,wald1975particle}. Remarkably, recent developments \cite{Penington2020,Almheiri2019,Almheiri2020a,2019arXiv191111977P} have shown that the inclusion of certain wormhole configurations in the gravitational path integral, macroscopically different than the original background geometry, lead to the conclusion that the Hawking radiation ``purifies itself'' after the so-called ``Page time'' defined as the time at which the coarse-grained entropy of the radiation equals the coarse-grained entropy of the black hole \cite{PhysRevLett.71.3743} which is given by the area of the horizon in Planck units \cite{bekenstein2020black,hawking1975particle}
\begin{align}
    S_{BH} = \frac{A}{4}.
\end{align}
This demonstrates that the late-time state is not thermal, restoring hope that predictability may survive in the backwards time direction. It is of course crucial to understand not only \textit{if} predictability is restored but \textit{when} and \textit{how} the information of the early-time quantum state escapes out of the black hole and into the radiation. 

A precise and operationally meaningful way to quantify this escape of information is to consider two microscopically distinct, but macroscopically indistinct evaporating black holes e.g.~two black holes formed from collapse of quantum matter in orthogonal quantum states with identical total conserved quantities.\footnote{These two states may be thought of as living in the same microcanonical energy band as we generally do not expect exact degeneracies in the spectrum.} If an observer has access to the radiation of one of the black holes, they may make a quantum measurement. If information has escaped from the black hole, then a judiciously chosen measurement will allow the observer to determine which of the two quantum states formed the black hole. The larger the amount of information that has escaped, the easier it is for the observer to distinguish the black holes using measurements on the radiation. To quantify this, we will study the \textit{quantum fidelity}, a useful measure of distinguishability. For two distinct evaporating black holes, Hawking's prediction corresponds to the fidelity equaling one at all times, meaning perfect indistinguishability. Our goal is to determine if, when, and how the fidelity can be found to be less than one, re-establishing predictability in quantum gravity (in simple toy models). 

The aforementioned progress on the entropy of Hawking radiation suggests that quantum information escapes the black hole only after the Page time. This is the time at which a region behind the horizon called the ``island'' forms, denoting the region whose quantum state is ``reconstructible'' on the radiation. This is consistent with the standard belief that there is no information in early Hawking radiation. The goal of this paper is to show that this belief receives important corrections and there is indeed genuine information regarding the black hole microstructure emanating within the radiation even before the Page time when no island is present. 

This work builds upon previous hints regarding nontrivial information in early Hawking radiation.\footnote{A complementary analysis of information in the Hawking radiation prior to the Page time was analyzed in \cite{PhysRevLett.129.061602, 2021arXiv211200020V} where it was argued that there is significant quantum entanglement in the radiation starting at a much earlier time scale referred to as $t_b$.} The computation of fidelity that we seek was first computed in \cite{2021PRXQ....2d0340K} for the PSSY model \cite{2019arXiv191111977P} involving two-dimensional Jackiw-Teitelboim gravity decorated with end-of-world branes that are entangled with an auxiliary ``radiation'' system. 
By summing over particular replica wormhole configurations in the Euclidean path integral, it was found that before the Page time, the fidelity was $F = 1 - \frac{1}{2}e^{-(S_{BH}-S_{rad})}+\dots$, which is very close to, though less than one, proving that in principle different black hole microstates may be distinguished prior to the Page time. Had replica wormholes not been included in the calculation, the fidelity would have been exactly one at all times, consistent with Hawking's analysis. The deviations from $F =1$ in this model were more systematically studied in the context in the JLMS formula \cite{2016JHEP...06..004J} in \cite{2022arXiv220311954K}. While this model is certainly illuminating, it has certain drawbacks such as not describing genuine time evolution of the evaporation process as well as lacking matter fields.

\begin{figure}
    \centering
    \includegraphics[width = \textwidth]{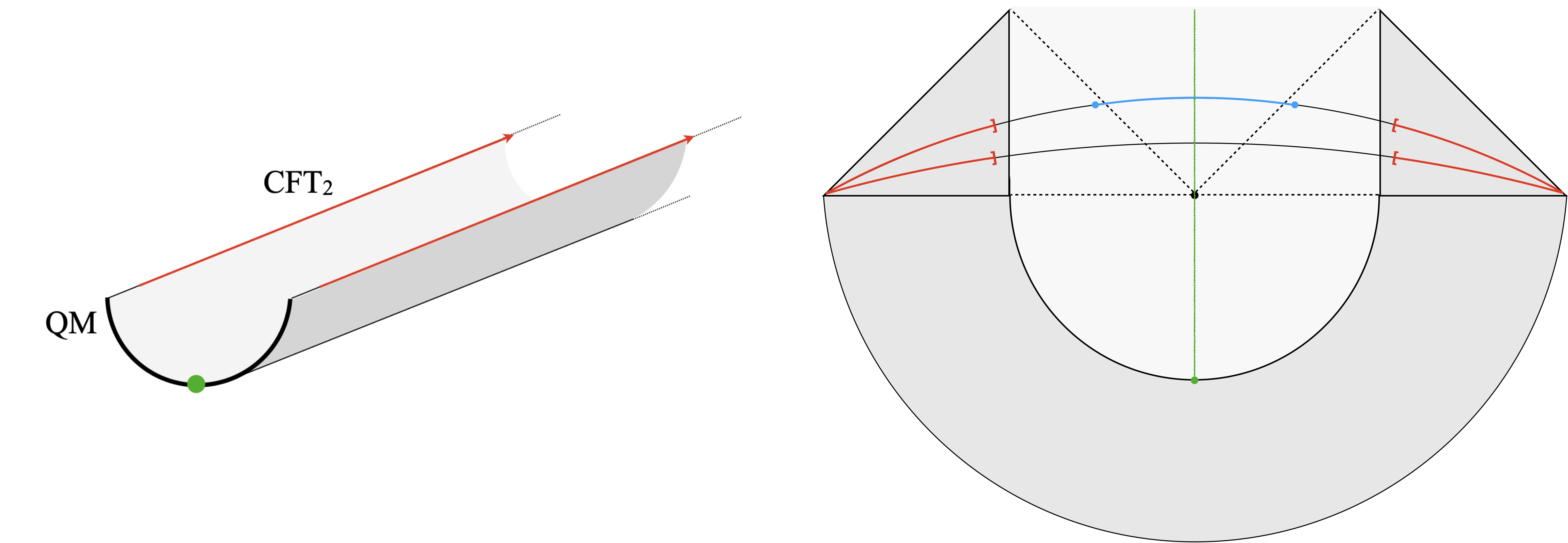}
    \caption{Left: the Euclidean path integral that prepares our state of interest. There are two copies of a boundary conformal field theory. The red, semi-infinite subregions account for the radiation while the black boundary together with cutoff region accounts for the black hole. An operator shown in green is inserted on the boundary. Right: the Lorentzian continuation of this state with the boundary degrees of freedom holographically represented as a 2D eternal black hole. The excitation intersects the bifurcation point and remains behind the horizon. After the Page time, an island (blue) forms that includes the bulk excitation.}
    \label{fig:setup}
\end{figure}

We are motivated to investigate more complicated evaporation models like those in \cite{2019arXiv191011077A, 2020JHEP...05..004R,Almheiri2020a}. These are built from a two dimensional conformal field theory whose boundary is coupled to a ``quantum dot'' or conformal boundary that is holographically dual to a two-dimensional gravitational system. Such a system can support non-trivial time evolution and a spectrum of matter fields. When prepared in the thermofield double state, the time evolution of this system may be interpreted as a two-dimensional black hole radiating into a bath\footnote{The term ``bath'' is a bit of a misnomer as it suggests Markovian dynamics which are not unitary by construction. We stick with tradition in calling this non-Markovian system a bath.} that is ``bulk'' conformal field theory. This conformal field theory thus comprises the ``radiation'' (see Figure \ref{fig:setup}).

The time evolution of the entanglement entropy (Page curve) for this model was considered from both the bulk and boundary perspectives in \cite{Almheiri2020, 2020JHEP...05..004R,2019arXiv191011077A} where it was found that at the Page time, an island forms that encompasses the black hole interior. We analyze the quantum fidelity between two distinct black holes in this model, modeled by placing particles of different flavors behind the black hole horizon, and find that there is significant information in the Hawking radiation prior to the formation of the island, with the fidelity between two evaporating black holes being
\begin{align}
    F = 1 -\pa{
C_{ppr}
-C_{qqr}
}^2
\frac{ \Gamma\pa{\frac{1}{2} + \Delta_r } }{ 4\sqrt{\pi} \Gamma\pa{1 + \Delta_r }  }
\ex{-2\Delta_r \pa{S_\text{BH}-S_{\text{rad}}}},
\end{align}
where $\Delta_r$ is the conformal dimension of the lightest primary field in the CFT spectrum and $C_{ppr}/C_{qqr}$ are the OPE coefficients between the operator placed behind the horizon and the lightest primary. As expected, this is not an identical answer to the PSSY model due to the increased complexity in the theory, though it shares the key qualitative feature that it is a finite amount less than one, albeit non-perturbatively suppressed in the black hole entropy, implying the escape of information prior to the Page time. We believe that this is a general feature of evaporating black holes. After the Page time, we find the fidelity to be nearly zero, implying that the black holes are easily distinguishable, a notion we make precise in the following section. We discuss the similarities and differences in the mechanisms leading to information leakage in the BCFT model and the PSSY model.

\paragraph{The Model}

% We first introduce the toy model of black hole evaporation that we will use. 
We consider a two-dimensional conformal field theory with a conformal boundary. If this CFT is ``holographic,'' the three-dimensional dual has the conformal boundary extend into the bulk geometry as a ``Cardy brane.'' We treat the two-dimensional theory of this brane as the black hole system while the remaining geometry may be considered the radiation. To provide nontrivial evaporation dynamics, we prepare the boundary conformal field theory in the thermofield double state
\begin{align}
    \ket{TFD} = \frac{1}{\sqrt{\mathcal{Z}(\beta)}}\sum_i e^{-\beta E_i/2} \ket{E_i}_1\otimes \ket{E_i}_2
\end{align}
and evolve with $H = H_1\otimes 1_2 + 1_1 \otimes H_2$
\begin{align}
    \ket{TFD(t)} = \frac{1}{\sqrt{\mathcal{Z}(\beta)}}\sum_i e^{-(\beta+4it) E_i/2} \ket{E_i}_1\otimes \ket{E_i}_2.
\end{align}
Under this time evolution, there will be a nontrivial Page curve. In order to add flavor to the black hole, we include a boundary excitation. More specifically, we act with an operator on the boundary that has been evolved by $\beta/4$ in imaginary time
\begin{align}
    \ket{TFD(t)^{(1)}} \propto \sum_i e^{-(\beta+4it) E_i/2} \mathcal{O}_1(0,-\beta/4) \ket{E_i}_1\otimes \ket{E_i}_2,
    \\
    \ket{TFD(t)^{(2)}} \propto \sum_i e^{-(\beta+4it) E_i/2} \mathcal{O}_2(0,-\beta/4) \ket{E_i}_1\otimes \ket{E_i}_2.
    \label{state_defs}
\end{align}
With this choice of imaginary time evolution, the gravitational picture includes a particle behind the horizon at all times.

\subsubsection*{Overview}

In Section \ref{sec:dist}, we provide background on the distinguishability measures that we compute, particularly the quantum fidelity. This makes precise how one should interpret our results in terms of an observers ability to know the details of the black hole formation from measurements only on the radiation. In Section \ref{entropy_sec}, we compute the entropy in the island model using conformal field theory techniques. In Section \ref{sec:fid}, we compute the fidelity in the island model, presenting the mechanisms leading to our main result and comparing these to the PSSY model. 
% In \secref{sec:circ}, we give an alternate perspective by studying a toy quantum circuit model, where similar phenomena arise.
While of a different general motivation, in Section \ref{sec:ETH}, we discuss how similar computations can be done to characterize eigenstate thermalization for extended subsystems in two-dimensional conformal field theory. 
Certain technical details are left to the appendices.

\section{What do distinguishability measures measure?}
\label{sec:dist}

In the midst of technical calculations, it is easy to lose sight of the concrete meanings of the quantities we are computing. We emphasize in this section the precise meaning of the quantum fidelity, which fortunately quantifies a very natural and concrete task that an experimentalist may perform.

The Uhlmann fidelity (which we will simply call \textit{the} fidelity) measures the distinguishability between two quantum states, $\rho$ and $\sigma$ and is defined as
\begin{align}
    F(\rho,\sigma) = \Tr \sqrt{\sqrt{\rho} \sigma\sqrt{\rho}} .
\end{align}
This quantity obeys several nice properties such as Jozsa's axioms \cite{Jozsa1994} and is bounded as
\begin{align}\label{eq:bound}
0\leq F(\rho,\sigma)\leq 1,
\end{align}
where the upper (lower) bound is saturated if and only if $\rho=\sigma$ ($\rho \sigma=0$).

A fundamental task in quantum information theory is that of \textit{quantum hypothesis testing}. Quantum hypothesis testing is the scenario where Alice is given a quantum state with the promise that it is either $\rho$ or $\sigma$. It is her task to make a quantum measurement to determine which state she was given. There is an error probability of her determining she has state $\rho$ when she really has state $\sigma$. Similarly, there is an error probability of Alice determining she has state $\sigma$ when she really has state $\rho$. The sum of these two error probability, $P_{err}$, for the optimized measurement\footnote{Such an optimized measurement can be explicitly constructed \cite{helstrom1969quantum}.} is bounded from both above and below by the fidelity as 
\begin{align}
    1-\sqrt{1-F(\rho,\sigma)^2} \leq P_{err} \leq {F(\rho,\sigma)}.
\end{align}
Moreover, if Alice is given $n$ copies of the state, she can make a more complicated measurement acting on all copies such that 
\begin{align}
    P_{err}^{(n)} \leq F(\rho,\sigma)^{n}.
    \label{eq:asyptotic_error}
\end{align}
Clearly, if the fidelity is less than one, if given a sufficient number of copies of the state, Alice may identify the state with high probability. This will be the case in the black hole setting where the sufficient number of copies will be exponentially large in the entropy of the black hole prior to the Page time. After the Page time, we find that only a single copy of the state is needed. 

The square roots present in the fidelity make it challenging (though still possible) to compute. Simpler quantities to analyze only involve the second moments of the density matrices. For illustration, we will consider the super-fidelity \cite{2008arXiv0805.2037M}
\begin{align}
    F_S(\rho,\sigma) := \Tr \rho \sigma + \sqrt{(1- \Tr \rho^2)(1- \Tr \sigma^2)} 
\end{align}
and the geometric mean (GM) fidelity \cite{2008PhLA..373...58W}
\begin{align}
    F_{GM}(\rho, \sigma) := \frac{\Tr \rho \sigma}{\sqrt{\Tr \rho^2 \Tr \sigma^2}}.
\end{align}
The super-fidelity is useful in that it upper bounds the fidelity. Therefore before the Page time, when we see that the super-fidelity is less than one, the fidelity must also be less than one. The GM fidelity is a useful proxy for the fidelity as it generally follows the same qualitative behavior even though it is neither an upper nor lower bound. This will be useful near the Page time where we do not have analytic control over the fidelity. It is also useful after the Page time because we find it to be non-zero which necessarily means that $\rho \neq \sigma$ so the fidelity itself must also be non-zero.

\section{Von Neumann Entropy}

\label{entropy_sec}

We warm up by considering the entropy of the radiation, deriving a Page curve. This calculation will also set our definition of the coarse-grained black hole and radiation entropies. The radiation system consists of the bulk conformal field theory from position $d$ to $\infty$ on both sides of the thermofield double (the red region in Figure \ref{fig:setup}), where $d$ is playing the role of a cutoff between the black hole and radiation subsystems. The coordinates, $z$, on the half cylinder have $\Re[z] \in [0,\infty)$ and $\Im[z] \in [0,2\pi)$.
% The thermofield double state with boundary may be prepared via a Euclidean path integral on the half cylinder.
We first conformally map the half cylinder to the disk using $w(z) = e^{\frac{2\pi}{\beta}z}$.
% \begin{align}\label{eq:state0}
%     \rho_R := \parbox{\boxrw}{\usebox{\boxr}}.
% \end{align}
% The right hand side of the equation is a representation of the path integral on the disk with an o that prepares $\rho_R$
The endpoints of the radiation subsystem are mapped to $w_1 =\bar{w}_1 =-e^{-\frac{2\pi}{\beta}d}$ and $w_2 =\bar{w}_2= e^{-\frac{2\pi}{\beta}d}$ on the disk. The moments of the density matrix may be computed using a two-point correlation function of twist operators
\begin{align}
\Tr\left( \rho_A^n\right)= 
\left({\frac{2\pi}{\beta}}\right)^{4h_n} 
\langle{ \sigma_n ( w_1, \bar{w}_1)   \bar{\sigma}_n (w_2, \bar{w}_2)   }\rangle_{\text{disk}}  .
\end{align}
We may evolve in imaginary time and subsequently analytically continue the correlation function to real time, such that
\begin{align}
w_1 = -\ex{-\frac{2\pi}{\beta}(d+t) },
\quad
w_2 = \ex{-\frac{2\pi}{\beta}(d-t) },
\quad
\bar{w}_1 = -\ex{-\frac{2\pi}{\beta}(-d+t)  },
\quad
\bar{w}_2 = \ex{-\frac{2\pi}{\beta}(-d-t) }.
\end{align}
For simplicity, we take the high-temperature limit ($\beta \rightarrow 0$). In this limit, the answer is independent of the specific details of the CFT. Later on, this limit will allow us to compute subleading corrections to the fidelity reliably. The correlation function on the disk can be naively ``unfolded'' to a correlation function without a boundary, an approximation that has been leveraged in the past to simplify replica calculations in boundary conformal field theory \cite{2015JHEP...09..110A,2020JHEP...04..074K}.
Due to the unfolding, the correlation function is defined on a chiral CFT.
The unfolding is not precise because we neglect a potential interface operator corresponding to the boundary.
This simplification replaces bulk-bulk-boundary OPE coefficients with the corresponding bulk-bulk-bulk OPE coefficients, though importantly does not change the qualitative behavior that we intend to isolate. One may prefer to call the brane in the bulk a $\mathbb{Z}_2$ brane or Hartman-Maldacena brane \cite{2013JHEP...05..014H} instead of a Cardy brane.

The high-temperature limit then corresponds to an OPE limit of the chiral twist fields such that at early times ($t<d$), when $w_1 \sim w_2$ and $\bar{w}_1 \sim \bar{w}_2$ 
\begin{align}
    \Tr\left( \rho_A^n\right)= \left({\frac{2\pi}{\beta}}\right)^{4h_n} (w_{1}-w_2)^{-2h_n}(\bar{w}_{1}-\bar{w}_2)^{-2h_n},
\end{align}
and at late times ($t>d$), when $w_1 \sim \bar{w}_1$ and $w_2 \sim \bar{w}_2$
\begin{align}
    \Tr\left( \rho_A^n\right)= \left({\frac{2\pi}{\beta}}\right)^{4h_n} (w_{1}-\bar{w}_1)^{-2h_n}({w}_{2}-\bar{w}_2)^{-2h_n}.
\end{align}
Both expressions may be analytically continued in $n$ to find the von Neumann entropy
\begin{align}
    S_{vN}(A) = \lim_{n\rightarrow 1}\frac{1}{1-n}\log \Tr \left( \rho_A^n\right)=
    \begin{cases}
    \frac{2\pi c}{3\beta} t + \frac{c}{3}\log \frac{\beta}{2\pi \epsilon}, & t < d
    \\
    \frac{2\pi c}{3\beta} d + \frac{c}{3}\log \frac{\beta}{2\pi \epsilon},& d < t
    \end{cases},
\end{align}
where $\epsilon$ is a UV cutoff.
There are subleading (in $\beta$) terms that also contribute including the boundary entropy, though we neglect these due to $\beta$ being our expansion parameter. In comparing to the Page curve, we can now identify $\frac{2\pi c}{3\beta} t$ as corresponding to the coarse-grained entropy of the radiation, $S_{rad}$, and $\frac{2\pi c}{3\beta} d$ as the coarse-grained entropy of the black hole, $S_{BH}$. A large boundary entropy may be added to the boundary but this will not significantly change the following formulas.

\section{Fidelity in the island model}
\label{sec:fid}

\subsection{Super and GM Fidelities}
\paragraph{PSSY Model}

To gain intuition, we first evaluate the super-fidelity and GM fidelity between two radiation states in the PSSY model. The analog of \eqref{state_defs} is
\begin{align}
    \ket{\Psi_1} &:= \frac{1}{\sqrt{k}}\sum_i^k \ket{\psi_{i,1}}_{B} \ket{i}_{R}, \quad \rho_R^{(1)} := \Tr_B \ket{\Psi_1}\bra{\Psi_1},
    \\
    \ket{\Psi_2} &:= \frac{1}{\sqrt{k}}\sum_i^k \ket{\psi_{i,2}}_{B} \ket{i}_{R},\quad \rho_R^{(2)} := \Tr_B \ket{\Psi_2}\bra{\Psi_2},
\end{align}
where, as described in \cite{2019arXiv191111977P}, $\ket{\psi_{i,a}}$ may be interpreted as a black hole state with the EOW brane in state $i$ and a small excitation with flavor $a$ propagating behind the horizon.

The purity of the radiation is the same for both states
\begin{align}
    \Tr \left(\rho_R^{(1)}\right)^2= \Tr \left(\rho_R^{(2)}\right)^2  = \frac{1}{k^2}\sum_{i,j} |\bra{\psi_{i,a}}\psi_{j,a}\rangle_B|^2, \quad a = 1,2.
\end{align}
Crucially, the inner product appearing in the sum is not proportional to a Kronecker delta. Instead, the inner product defines the following boundary conditions in the gravitational path integral
\begin{align}
   \includegraphics[height = 2.8cm]{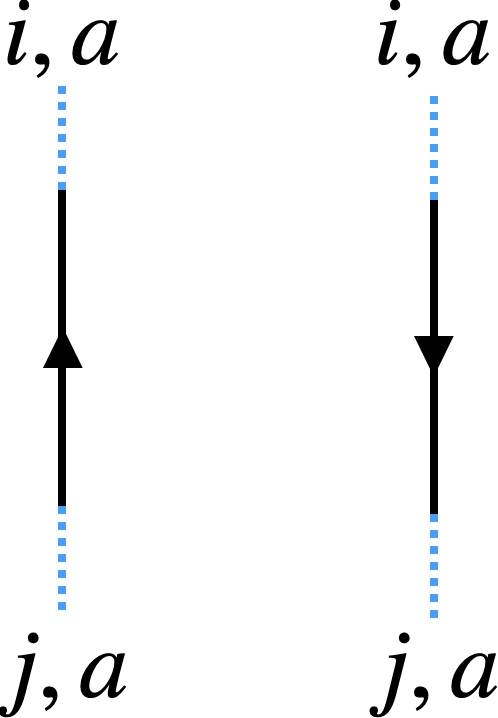}
\end{align}
where the black lines represent the asymptotic boundaries and the dotted blue lines enforce that an EOW brane with a particular flavor intersects the boundary. These boundary conditions can be filled in in two topologically distinct ways,\footnote{We ignore higher genus solutions because these are exponentially suppressed in the large ground state entropy $S_0$.} corresponding to a factorized bulk path integral and a wormhole solution
\begin{align}
    \includegraphics[height = 4cm]{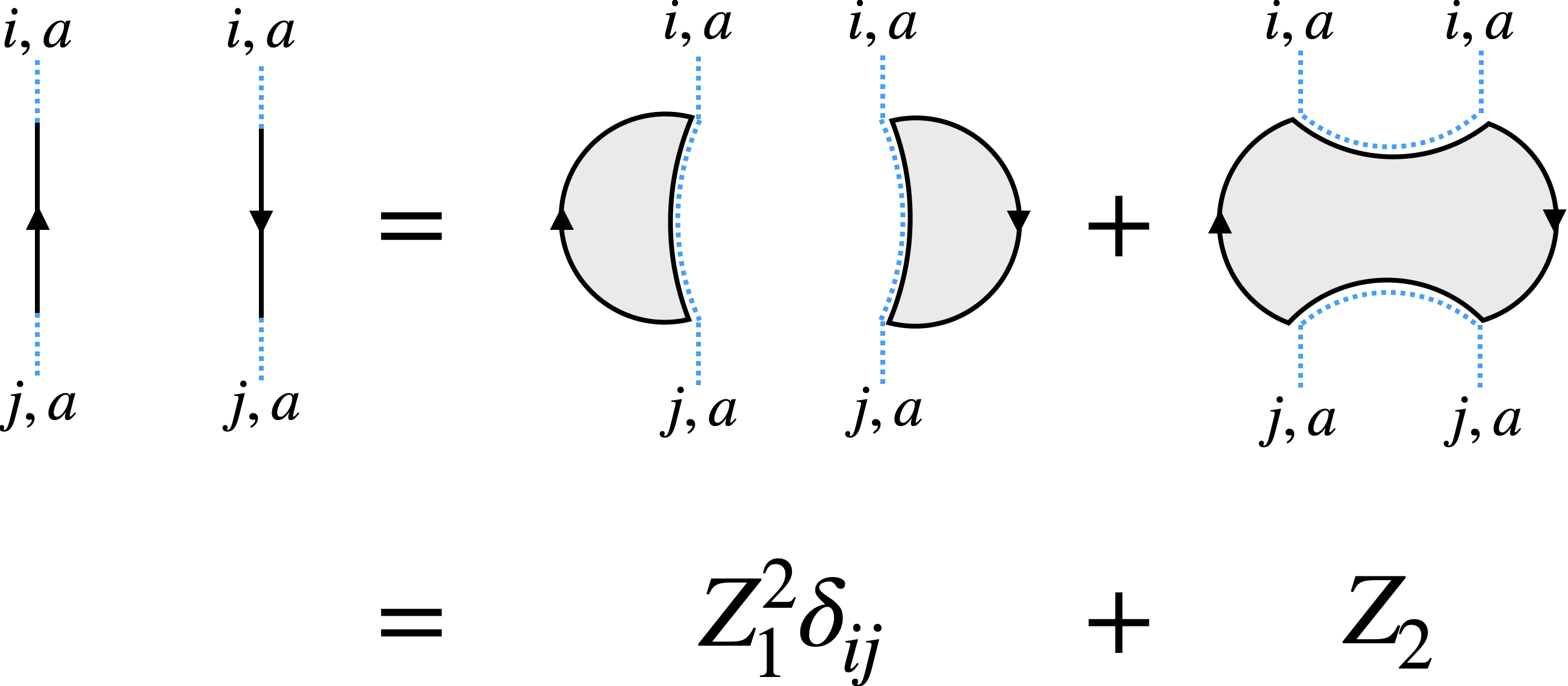}
    \label{purity_pssy_cartoon_eq}
\end{align}
Completing the sums over $i$ and $j$ and including the normalization of the state, $Z_1^{-2}$, we find
\begin{align}
    \Tr \left(\rho_R^{(1)}\right)^2= \Tr \left(\rho_R^{(1)}\right)^2  = \frac{1}{k} + \frac{Z_2}{Z_1^2}.
\end{align}
We interpret the first term as $e^{-S_{rad}}$ and the second term as $e^{-S_B^{(2)}}$ where $S_B^{(n)}$ is the coarse-grained R\'enyi entropy of the black hole.

\begin{figure}
    \centering
    \includegraphics[height = 4cm]{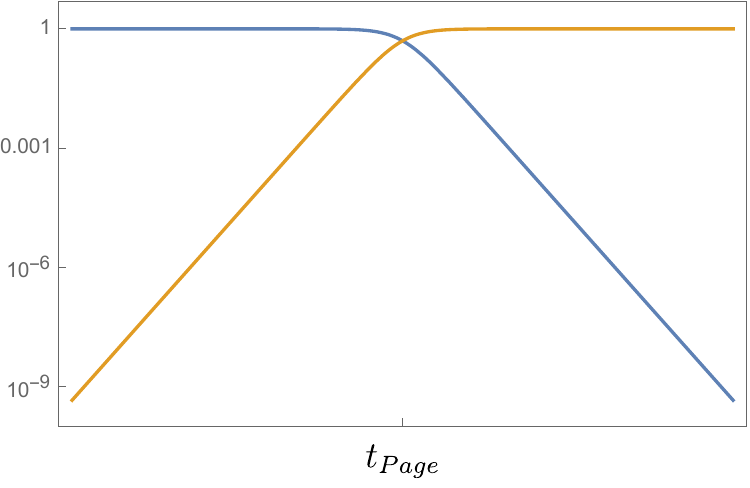}
    \includegraphics[height = 4cm]{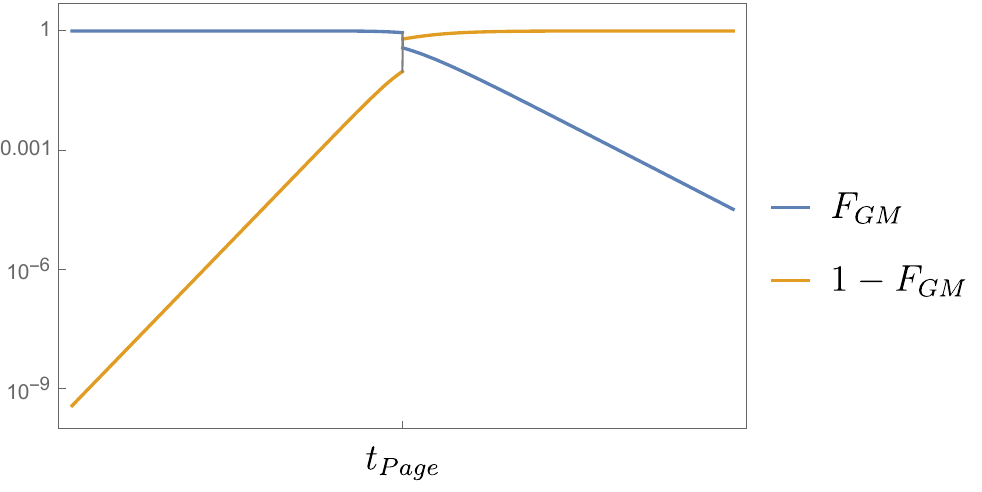}
    \caption{The GM fidelity in the PSSY (left) and BCFT (right) models.}
    \label{GM_fid_bcft}
\end{figure}

The overlap is given by
\begin{align}
    \Tr \left(\rho_R^{(1)} \rho_R^{(2)}\right)  = \frac{1}{k^2}\sum_{i,j} \bra{\psi_{i,1}}\psi_{j,1}\rangle_B\bra{\psi_{i,2}}\psi_{j,2}\rangle_B,
\end{align}
which sets the boundary conditions as 
\begin{align}
    \includegraphics[height = 4cm]{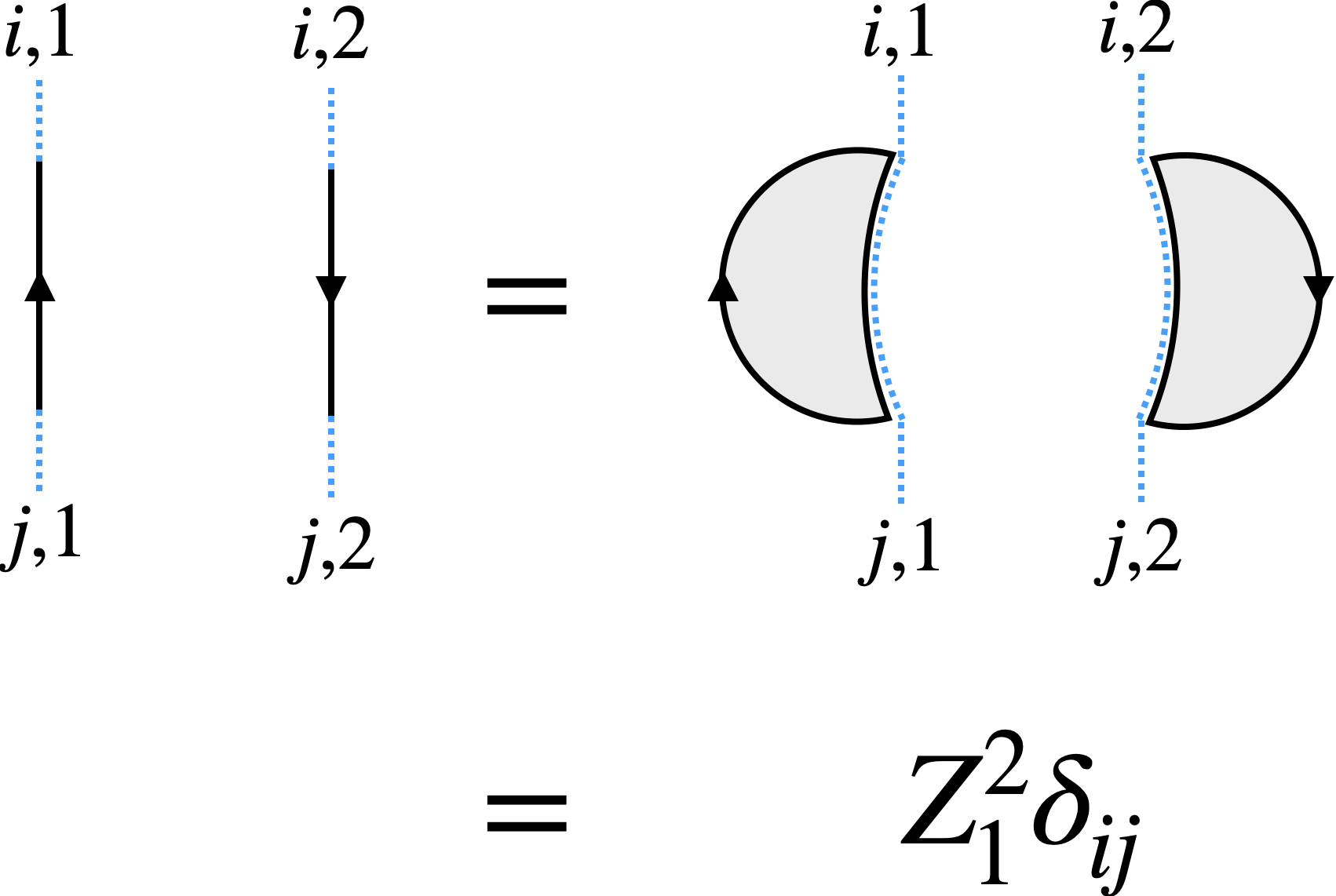}
    \label{overlap_pssy_cartoon_eq}
\end{align}
Now, only the disconnected solution contributes to the path integral because the brane indices are incompatible, not allowing a wormhole. Summing over $i$ and $j$, we find 
\begin{align}
    \Tr \left(\rho_R^{(1)} \rho_R^{(2)}\right)  = e^{-S_{rad}}.
\end{align}
The super-fidelity and GM fidelity are subsequently
\begin{align}
    F_S(\rho_R^{(1)} \rho_R^{(2)}) = 1-e^{-S^{(2)}_{B}} , \quad F_{GM}(\rho_R^{(1)} \rho_R^{(2)})  = \frac{1}{{1+e^{-(S^{(2)}_{B}-S_{rad})}}}.
\end{align}
The super-fidelity implies that the fidelity is less than one before the Page time by an amount at least exponentially small in the black hole entropy. This implies that an observer may distinguish the two black holes given an exponentially large number copies of the state of the radiation. The GM fidelity displays a Page curve that transitions between close to one and close to zero at the Page time (see Figure \ref{GM_fid_bcft}). Using heavier machinery involving higher moments, one finds that the GM fidelity faithfully mimics the behavior seen in the fidelity \cite{2021PRXQ....2d0340K}.

\paragraph{The BCFT Model}

We now demonstrate the analogs of these calculations in the BCFT model by considering the operators in \eqref{state_defs} to have conformal dimensions $1 \ll \Delta \ll c$ and the CFT to be ``holographic.'' Later, we will provide more general calculations. In this regime of dimensions, the operators behave as generalized free fields and their correlations functions may be computed using Wick contractions. Moreover, their bulk duals are quantum fields that are massive enough to follow classical trajectories but light enough as to not backreact on the geometry.

The Wick contractions between different replicas are somewhat analogous to the replica wormhole contributions from \eqref{purity_pssy_cartoon_eq}. Because $\phi_p$ and $\phi_q$ are orthogonal fields (their two-point function is trivial), their contractions do not contribute to the path integral. This mimics the phenomenon where replica wormholes in \eqref{overlap_pssy_cartoon_eq} connecting different flavored EOW branes do not contribute to the gravitational path integral.

Using the replica path integral approach, the purities and overlaps of the two states can be seen to be given by the following correlation functions
\begin{align}
\label{sigma22_eq}
   \Tr \left(\rho_R^{(a)}\rho_R^{(b)}\right) = \langle{X^{\dagger}_{ab}(i) X_{ab}(-i) }\rangle_{ \Sigma_{2,2}  },
\quad
X_{ab} = \mathcal{O}_a\otimes \mathcal{O}_b,
\end{align}
where $\Sigma_{2,2}$ is the manifold with two replica sheets glued along two intervals with endpoints (after analytic continuation to Lorentzian time) at
\begin{align}
z_1 = -\ex{-\frac{2\pi}{\beta}(d+t) },
\quad
z_2 = \ex{-\frac{2\pi}{\beta}(d-t) },
\quad
z_3 = \ex{-\frac{2\pi}{\beta}(-d-t) },
\quad
z_4 = -\ex{-\frac{2\pi}{\beta}(-d+t)  }.
\end{align}
The operator $X_{ab}$ is bi-local, being a tensor product of primary operators located at the same point but on separate sheets. This should not be confused with an operator in the orbifolded theory.
This manifold has the topology of a torus. For the purity ($a = b$), there are two Wick contractions, one contracting operators on the same sheet and the other contracting operators on opposite sheets. While the difference in magnitude of these terms is large, it is important to include both in order to find the leading corrections to the fidelity. For the overlap ($a \neq b$), there is only a single Wick contraction, just as in the PSSY model.

Famously, the torus partition function undergoes a first-order phase transition when the two cycles have equal length, corresponding in gravity to the Hawking-Page transition \cite{hawking1983thermodynamics}. In our model this signals the Page transition and there is a change in the functional form of the Wick contractions. The precise form of the GM and super fidelities are somewhat complicated and can be found in Appendix \ref{detail_GFF}. Here, we emphasize the fact that they mimic the results from the PSSY model in that they imply that the fidelity is less than one by an amount exponentially suppressed by the black hole entropy before the Page time and are non-zero but very small after the Page time. We plot the GM fidelity in Figure \ref{GM_fid_bcft}, comparing the curve to the simple answer from the PSSY model. The fidelities are seen to have very similar behavior.

\subsection{Uhlmann fidelity in the BCFT Model}
\label{uhl_subsec}

To compute the full Uhlmann fidelity, we need to compute higher moments of the density matrices by using a replica trick \cite{Lashkari2014, Zhang2019, Kusuki2020},
\begin{align}
F(\rho,\sigma)=  \lim_{m,n \rightarrow  \frac{1}{2}}\Tr \left({\rho^m \sigma \rho^m}\right)^n  .
\end{align}
To leading order, the fidelity was computed between two states excited above the vacuum with heavy operators \cite{Kusuki2020}. Using the same techniques, we may compute the leading order Page curve for the fidelity. This will simply lead to a step function with the fidelity equalling one before the Page time and zero after the Page time. The subleading corrections to this calculation are difficult to control in generality, so we instead use a separate method in the following section.

The moments may be evaluated using a $2k$-point function on, $\Sigma_{k,2}$, the $k$-sheeted replica manifold glued along two intervals
\begin{align}
\label{replica_corr_eq}
\begin{aligned}
    \Tr \left({\left(\rho_R^{(1)}\right)^m \rho_R^{(2)}\left( \rho_R^{(1)}\right)^m}\right)^n &= \frac{\langle X_{k}^{\dagger}(i) X_{k}(-i)\rangle_{\Sigma_{k,2}}}{\langle \phi_p^{\dagger}(i) \phi_p(-i)\rangle_{\mathbb{C}}^{2mn}\langle \phi_q^{\dagger}(i) \phi_q(-i)\rangle_{\mathbb{C}}^{n}}, 
    \\
    X_{k} &= \left(\phi_p^{\otimes m} \otimes \phi_q \otimes \phi_p^{\otimes m}\right)^{\otimes n}.
\end{aligned}
\end{align}
where $k := (2m+1)n$. 
Using the Riemann-Hurwitz formula, one can see that the replica manifold is a $(k-1)$-genus surface and thus extremely difficult to evaluate in generality. We only consider the leading Wick contraction. Prior to the Page time, this contraction connects operators on the same replica sheet which is identical to the normalization factor, so the fidelity is trivially one. After the Page time, the Wick contraction of the closest pairs of operators connects operators on cyclically permuted sheets. Because these involve contracting $\phi_p$ with $\phi_q$, we find that the fidelity is zero. To find a non-zero contribution, one needs sum contractions involving distant operators. The denominator will then dominate significantly over this contribution giving a very small fidelity. We already understood from the previous subsection that the fidelity is not exactly zero because otherwise the GM fidelity would be exactly zero. We now characterize this leading contribution.
    
\paragraph{The high-temperature limit}

We proceed to a more general calculation that may be done in the high-temperature ($\beta\rightarrow 0$) limit. The following calculations are more general in that they are do not place requirements on the dimensions of the boundary primary operators and the CFT is unconstrained. For example, the CFT can be a free boson.

It is convenient to describe the replica partition function using twist operators on the complex plane rather than a correlation function of primary fields on the $(k-1)$-genus surface
\begin{align}\label{eq:Zdisk}
\begin{aligned}
Z_{n,m}&=
\langle{X_k(i) \sigma_k(z_1)  \overline{\sigma}_k(z_2)  X_k(-i) }\rangle_{\text{disk}, \mathcal{C}^{\otimes k}/\mathbb{Z}_k},
\\
X_k &:= \pa{\phi_p^{\otimes m} \otimes \phi_q \otimes  \phi_p^{\otimes m}}^{\otimes n},
\end{aligned}
\end{align}
This expression is purely formal because the operator $X_k$ is not included in the spectrum of the orbifold theory $\mathcal{C}^{\otimes k}/\mathbb{Z}_k$ because $X_k$ is not symmetric under cyclic permutations.
Nevertheless, this notation is useful as explained in \cite{Sarosi2016}. This is because we can take the OPEs of operators on each sheet in \eqref{replica_corr_eq}. Grouping like-terms among the sheets, we find operators that are indeed symmetric under cyclic permutation, and under orbifolding reside in the untwisted sector. The effective OPE coefficients with an operator $T$ in the untwisted sector are dependent on how we take the OPE. We will write these explicitly shortly.

Again, we avoid the technical complications of the conformal boundary without discarding any of the essential physics by using the doubling trick and neglecting the contribution from the potential interface operator
\begin{align}
Z_{n,m}&=\langle{X_k(i) \sigma_k(z_1)  \overline{\sigma}_k(z_2) \sigma_k(z_3)  \overline{\sigma}_k(z_4)  X_k(-i) }\rangle_{ \mathcal{C}^{\otimes k}/\mathbb{Z}_k  }, 
\end{align}
where the insertion points are
\begin{align}
z_1 = -\ex{-\frac{2\pi}{\beta}(d+t) },
\quad
z_2 = \ex{-\frac{2\pi}{\beta}(d-t) },
\quad
z_3 = \ex{-\frac{2\pi}{\beta}(-d-t) },
\quad
z_4 = -\ex{-\frac{2\pi}{\beta}(-d+t)  }.
\end{align}

%%%%%%%%%%%%%%%%%%%%%%%%%%%%%%%%%%%%%%%%%%%%%%%%%%%%%%%%%%%%%%%%%%%%%%%%%%%%%%%%%%%%%%%%%%%%%%
\paragraph{Before the Page time}
%%%%%%%%%%%%%%%%%%%%%%%%%%%%%%%%%%%%%%%%%%%%%%%%%%%%%%%%%%%%%%%%%%%%%%%%%%%%%%%%%%%%%%%%%%%%%%
Let us evaluate the correlation function before Page time.
For convenience, we apply the following conformal map,
\begin{align}
w = \frac{z+i}{z-i}.
\end{align}
The correlation function is transformed to
\begin{align}
Z_{n,m} = \text{(conformal factor)}\langle{X_k(0) \sigma_k(w_1)  \overline{\sigma}_k(w_2) \sigma_k(w_3)  \overline{\sigma}_k(w_4)  X_k(\infty) }\rangle_{ \mathcal{C}^{\otimes k}/\mathbb{Z}_k  }.
\end{align}
We leave the conformal factor implicit because the factors corresponding to the local operators cancel via the denominator of \eqref{replica_corr_eq} and the factors corresponding to the twist fields are trivial in $k \rightarrow 1$ limit because the dimensions of the twist fields go to zero. 
In the high-temperature limit, the insertion points can be approximated at early times ($t<d$) as
\begin{align}
\begin{aligned}
w_1 &= -1 -2i\ex{-\frac{2\pi}{\beta}(d+t) },
\quad
w_2 = -1+2i\ex{-\frac{2\pi}{\beta}(d-t) },
\\
w_3 &= 1+2i\ex{-\frac{2\pi}{\beta}(d+t) },
\quad \hspace{.3cm}
w_4 =  1-2i \ex{-\frac{2\pi}{\beta}(d-t)  }.
\end{aligned}
\end{align}
We expand the correlation function in this limit, taking the OPEs of pairs of operators on each sheet
\begin{align}
Z_{n,m} = \pa{2\ex{-\frac{2\pi}{\beta}\pa{d-t}}}^{-4h_k}
\pa{
1+ { 2}
\sum_{T \in \mathcal{S}}
\hat{C}_{X_k X_k T} \hat{C}_{\sigma_k \bar{\sigma}_k T} (-1)^{\frac{l_T}{2}}  \pa{2 \ex{-\frac{2\pi}{\beta}(d-t)}}^{\Delta_T} + \cdots
},
\end{align}
where $\mathcal{S}$ is the set of the lightest non-vacuum fields. We denote the scaling dimension by $\Delta$ and the spin by $l$.
Any field in the untwisted sector of the orbifold CFT can be described as
\begin{align}
T:= \frac{1}{k} \pa{T_0\otimes \cdots T_{k-1} + (\text{cyclic permutations})  }.
\end{align}
For these untwisted states,
we define the OPE coefficients by
\begin{align}
\hat{C}_{\sigma_k \bar{\sigma}_k T} := \langle{\sigma_k(0)  T(1)  \overline{\sigma}_k(\infty) }\rangle_{\mathcal{C}^{\otimes k}/\mathbb{Z}_k},
\end{align}
and
\begin{align}
\begin{aligned}
\hat{C}_{X_k X_k T} 
&:=
\prod_{k=0}^{n-1}
\pa{\prod_{i=0}^{m-1} C_{ppT_{i+(2m+1)k}}C_{ppT_{i+m+(2m+1)k}}}
C_{qqT_{m+(2m+1)k}}
% \\&
+ (\text{cyclic permutations}).
\end{aligned}
\end{align}
The lightest non-vacuum contributions have the following form, 
\begin{align}\label{eq:first}
\Psi^i=\frac{1}{k} \pa{ \psi_r \otimes \mathbb{I}^{\otimes i}  \otimes \psi_r \otimes \mathbb{I}^{\otimes k-i-1}   + (\text{cyclic permutations}) },
\quad i \leq \frac{k}{2}.
\end{align}
For these states, the OPE coefficients may be evaluated as
\begin{align}\label{eq:sum_OPE}
\begin{aligned}
&
\sum_{T \in \mathcal{S}}
\hat{C}_{X_k X_k T} \hat{C}_{\sigma_k \bar{\sigma}_k T} 
(-1)^{\frac{l_T}{2}} 
\\
&=
\sum_{\substack{i=1\\(2m+1) \nmid i}}^{k/2}
\pa{
n(2m-1)C_{ppr}^2 + 2n C_{ppr}  C_{qqr}
}
C_{\sigma_k \bar{\sigma}_k \Psi^i_r }
(-1)^{l_r} 
% \\
% &
+
\sum_{\substack{i=1 \\ (2m+1)|i}}^{k/2}
\pa{
2nmC_{ppr}^2
+nC_{qqr}^2
}
C_{\sigma_k \bar{\sigma}_k \Psi^i_r }
(-1)^{l_r} 
\\
&=
\sum_{\substack{i=1}}^{k/2}
\pa{
n(2m-1)C_{ppr}^2+ 2n  C_{ppr} C_{qqr}
}
C_{\sigma_k \bar{\sigma}_k \Psi^i_r }
(-1)^{l_r} 
% \\
% &
+
\sum_{\substack{i=1 \\ (2m+1)|i}}^{k/2}
n
\pa{
C_{ppr}
-C_{qqr}
}^2
C_{\sigma_k \bar{\sigma}_k \Psi^i_r }
(-1)^{l_r} 
\end{aligned}
\end{align}
We first take the $m\rightarrow 1/2$ limit
\begin{align}
 2n C_{ppr} C_{qqr}
\sum_{i=1}^{2n-1} \frac{1}{\pa{4n\sin \frac{\pi i}{2n}}^{2\Delta_r}}
% \\
% &
+
\frac{n}{2}
\pa{
C_{ppr}
-C_{qqr}
}^2
\sum_{i=1}^{n-1} \frac{1}{\pa{4n\sin \frac{\pi i}{n}}^{2\Delta_r}}
\end{align}
where we use the relation \cite{Headrick2010},\footnote{In \cite{Headrick2010}, there is an additional factor $n$,
which comes from the cyclic permutation.
We do not need this factor here because we already included all terms of the cyclic permutation in the summand of (\ref{eq:sum_OPE}).
}
\begin{align}
\begin{aligned}
(-1)^{l_r }
\sum_{\substack{i=1}}^{k/2} C_{\sigma_k \bar{\sigma}_k \Psi^i_r }
= \frac{1}{2} \sum_{i=1}^{k-1} \frac{1}{\pa{2k\sin \frac{\pi i}{k}}^{2\Delta_r}} := \frac{1}{2}f(k,\Delta_r).
\end{aligned}
\end{align}
We then take the $n\rightarrow 1/2$ limit
\begin{align}
\frac{1}{2^{2\Delta_r+2}}
\pa{
C_{ppr}
-C_{qqr}
}^2
f\pa{\frac{1}{2},\Delta_r}.
\end{align}
% We use the fact \cite{Calabrese2011},
% \begin{align}\label{eq:f=0}
% f(1,h)=0.
% \end{align}
We conjecture using numerical calculations\footnote{In practice, this means evaluating $f(n,h)$ for fixed integer $h$ and general $n$, identifying (using Mathematica) an analytic function of $n$ that reproduces the sequence at fixed $h$, analytically continuing to $n = 1/2$, then identifying an analytic function that reproduces the sequence in $h$.} that the analytic continuation $n\to \frac{1}{2}$ is given by\footnote{We would like to thank Nathan Benjamin for informing us of this conjectured expression.}
\begin{align}
f\pa{\frac{1}{2},h}=- \frac{ \Gamma\pa{\frac{1}{2} + h } }{ 2\sqrt{\pi} \Gamma\pa{1 + h }  } < 0.
\end{align}
This is consistent with all explicitly summable examples, such as $h = 0, 1$.
% For example, if one considers the case where $\psi_r$ is a current ($\Delta_r=1$),
% the sum gives
% \begin{align}
% f\pa{n,1}=\frac{n^2 -1}{12n^2} 
% \Rightarrow
% f\pa{\frac{1}{2},1}=-\frac{1}{4}.
% \end{align}
% Other known examples are
% \begin{align}
% f\pa{\frac{1}{2},\frac{1}{2}}=-\frac{1}{\pi}, \quad 
% f\pa{\frac{1}{2},0}=-\frac{1}{2}.
% \end{align}
Crucially, this function is everywhere negative. For consistency, this had to be the case because the fidelity is bounded above by one.
Finally, we obtain
\begin{align}
\label{fid_final_eq}
F(\rho^p_A,\rho^q_A) = 1
-
\pa{
C_{ppr}
-C_{qqr}
}^2
\frac{ \Gamma\pa{\frac{1}{2} + \Delta_r } }{ 4\sqrt{\pi} \Gamma\pa{1 + \Delta_r }  }
\ex{-2\Delta_r \pa{S_\text{BH}-S_{\text{rad}}}}
.
\end{align}
For descendants of the vacuum, the OPE coefficients for the two primary fields are identical
\begin{align}
C_{ppr}
=C_{qqr},
\end{align}
where we use the assumption $h_p=h_q$.
Therefore, the first nontrivial contribution is given by the first non-vacuum \textit{primary} state.

%%%%%%%%%%%%%%%%%%%%%%%%%%%%%%%%%%%%%%%%%%%%%%%%%%%%%%%%%%%%%%%%%%%%%%%%%%%%%%%%%%%%%%%%%%%%%%
% \subsubsection{After Page time}
%%%%%%%%%%%%%%%%%%%%%%%%%%%%%%%%%%%%%%%%%%%%%%%%%%%%%%%%%%%%%%%%%%%%%%%%%%%%%%%%%%%%%%%%%%%%%%
\paragraph{After the Page time}
After Page time ($t>d$), the insertion points can be approximated as
\begin{align}
\begin{aligned}
w_1 &= -1 -2i\ex{-\frac{2\pi}{\beta}(d+t) },
\quad
w_2 = 1+2i\ex{-\frac{2\pi}{\beta}(t-d) },
\\
w_3 &= 1+2i\ex{-\frac{2\pi}{\beta}(d+t) },
\quad \hspace{.3cm}
w_4 =  -1-2i \ex{-\frac{2\pi}{\beta}(t-d)  }.
\end{aligned}
\end{align}
In this case, the high-temperature limit corresponds to operators on neighboring replica sheets to become close, so we wind up with a different OPE limit. The correlation function is correspondingly modified to
\begin{align}
\langle{X_k(i) \sigma_k(z_1)  \overline{\sigma}_k(z_2) \sigma_k(z_3)  \overline{\sigma}_k(z_4)  X'_k(-i) }\rangle_{ \mathcal{C}^{\otimes k}/\mathbb{Z}_k  }, 
\quad X'_k := \pa{\phi_p^{\otimes (m+1)} \otimes \phi_q \otimes  \phi_p^{\otimes (m-1)}}^{\otimes n}.
\end{align}
We can expand the correlation function in the high-temperature limit as
\begin{align}
\pa{2\ex{-\frac{2\pi}{\beta}\pa{t-d}}}^{-4h_k}
\pa{
1+2
\sum_{T \in \mathcal{S}}
\hat{C}_{X_k X_k T} \hat{C}_{\sigma_k \bar{\sigma}_k T} (-1)^{\frac{l_T}{2}}  \pa{2 \ex{-\frac{2\pi}{\beta}(t-d)}}^{\Delta_T} + \cdots
},
\end{align}
where $\mathcal{S}$ is a set of the first non-vacuum states.
The OPE coefficient is given by
\begin{align}
\begin{aligned}
\hat{C}_{X_k X'_k T}
&:=
\prod_{k=0}^{n-1}
\pa{\prod_{i=0}^{m-1} C_{ppT_{i+ (2m+1)k}}C_{ppT_{i+m+ (2m+1)k}}}
C_{pqT_{m+ (2m+1)k}}
C_{pqT_{m+1+ (2m+1)k}}\\
&+ (\text{cyclic pernutations}).
\end{aligned}
\end{align}
The first non-vacuum contribution is again given by (\ref{eq:first}).
Consequently, we obtain
\begin{align}
\begin{aligned}
&
\sum_{T \in \mathcal{S}}
\hat{C}_{X_k X'_k T} \hat{C}_{\sigma_k \bar{\sigma}_k T} 
(-1)^{\frac{l_T}{2}} 
% \\
% &
=\sum_{\substack{i=1\\(2m+1) \nmid (i+l) \\ l=-1,0,1}}^{k/2}
\pa{
n(2m-3)C_{ppr}^2 + 4n C_{ppr}  C_{pqr} 
}
C_{\sigma_k \bar{\sigma}_k \Psi^i_r }
(-1)^{l_r} 
\\
&+
\sum_{\substack{i=1 \\ (2m+1)|i}}^{k/2}
\pa{
n(2m-1)C_{ppr}^2
+2nC_{pqr}^2
}
C_{\sigma_k \bar{\sigma}_k \Psi^i_r }
(-1)^{l_r} 
\\
 &
+
2\sum_{\substack{i=1 \\ (2m+1)|(i+1)}}^{k/2}
\pa{
n(2m-2)C_{ppr}^2
+2n C_{ppr}C_{pqr}
+nC_{pqr}^2
}
C_{\sigma_k \bar{\sigma}_k \Psi^i_r }
(-1)^{l_r} 
\\
&=\sum_{\substack{i=1}}^{k/2}
\pa{
n(2m-3)C_{ppr}^2 + 4n C_{ppr} C_{pqr} 
}
C_{\sigma_k \bar{\sigma}_k \Psi^i_r }
(-1)^{l_r} 
% \\
% &
+\sum_{\substack{i=1 \\ (2m+1)|i}}^{k/2}
2n
\pa{
C_{ppr} - C_{pqr}
}^2
C_{\sigma_k \bar{\sigma}_k \Psi^i_r }
(-1)^{l_r} 
\\
&
+
\sum_{\substack{i=1 \\ (2m+1)|(i+1)}}^{k/2}
2n
\pa{
C_{ppr} - C_{pqr}
}^2
C_{\sigma_k \bar{\sigma}_k \Psi^i_r }
(-1)^{l_r} 
% \\
% &
\end{aligned}
\end{align}
In the $m\rightarrow 1/2$ limit, this equals
\begin{align}
    % &\ar{m \to \frac{1}{2}}
\pa{
-2nC_{ppr}^2 + 4n C_{ppr}C_{pqr} 
}
\sum_{i=1}^{2n-1} \frac{1}{\pa{4n\sin \frac{\pi i}{2n}}^{2\Delta_r}}
+
n
\pa{
C_{ppr} - C_{pqr}
}^2
\sum_{i=1}^{2n-1} \frac{1}{\pa{4n\sin \frac{\pi i}{2n}}^{2\Delta_r}}
\end{align}
which disappears when taking $n\rightarrow 1/2$, so the fidelity vanishes. We find this to be a somewhat miraculous cancellation, given the complexity of the replica calculation. Of course, we know that the fidelity cannot be exactly zero due to our calculation of the GM fidelity. However, this calculation shows that the non-zero contributions must be more subleading than $O\pa{\ex{-2\Delta_r \pa{S_{\text{rad}} -S_\text{BH} }}}$.
% Here we again use the fact (\ref{eq:f=0}).
% Again, our result is consistent with  $0\leq F(\rho^p_A,\rho^q_A)\leq 1$.

%%%%%%%%%%%%%%%%%%%%%%%%%%%%%%%%%%%%%%%%%%%%%%%%%%%%%%%%%%%%%%%%%%%%%%%%%%%%%%%%%%%%%%%%%%%%%%
% \subsection{Summary}
%%%%%%%%%%%%%%%%%%%%%%%%%%%%%%%%%%%%%%%%%%%%%%%%%%%%%%%%%%%%%%%%%%%%%%%%%%%%%%%%%%%%%%%%%%%%%%
We have found that before the Page time, the fidelity is close to one \eqref{fid_final_eq}.
% \begin{align}
% F(\rho^p_A,\rho^q_A) = 1
% -
% \pa{
% C_{ppr}
% -C_{qqr}
% }^2
% \frac{ \Gamma\pa{\frac{1}{2} + \Delta_r } }{ 4\sqrt{\pi} \Gamma\pa{1 + \Delta_r }  }
% \ex{-2\Delta_r \pa{S_\text{BH}-S_{\text{rad}}}}
% .
% \end{align}
Using \eqref{eq:asyptotic_error}, we see that ones needs an $O(e^{\#(S_{BH}-S_{rad}}))$ number of copies of the state of the radiation prior to the Page time in order to distinguish microstates. This is qualitatively the same conclusion as in \cite{2021PRXQ....2d0340K} in the simpler PSSY model. Moreover, the lighter the operators in the spectrum of the given CFT, the more easily the states can be distinguished. The light bulk fields appear to mediate the transfer of information from inside the black hole into the radiation system. This is a new feature which has no analog in \cite{2021PRXQ....2d0340K}.

After the Page time, the fidelity is close to zero
\begin{align}
F(\rho^p_A,\rho^q_A) =
o\pa{\ex{-2\Delta_r \pa{S_{\text{rad}} -S_\text{BH} }}}.
\end{align}
Therefore, a single judiciously chosen measurement of the radiation distinguishes the two different states of the black hole. While we have demonstrated that these measurements are effective, we have not shown that they are feasible in practice. Indeed one may expect that they are exponentially complex. While a disjoint concept from the unitarity of black hole evaporation, it would be interesting to understand if and when simpler measurements of the radiation may distinguish black holes.

\section{Fidelity between primary states}
\label{sec:ETH}

\begin{figure}
 \begin{center}
  \includegraphics[width=4.0cm,clip]{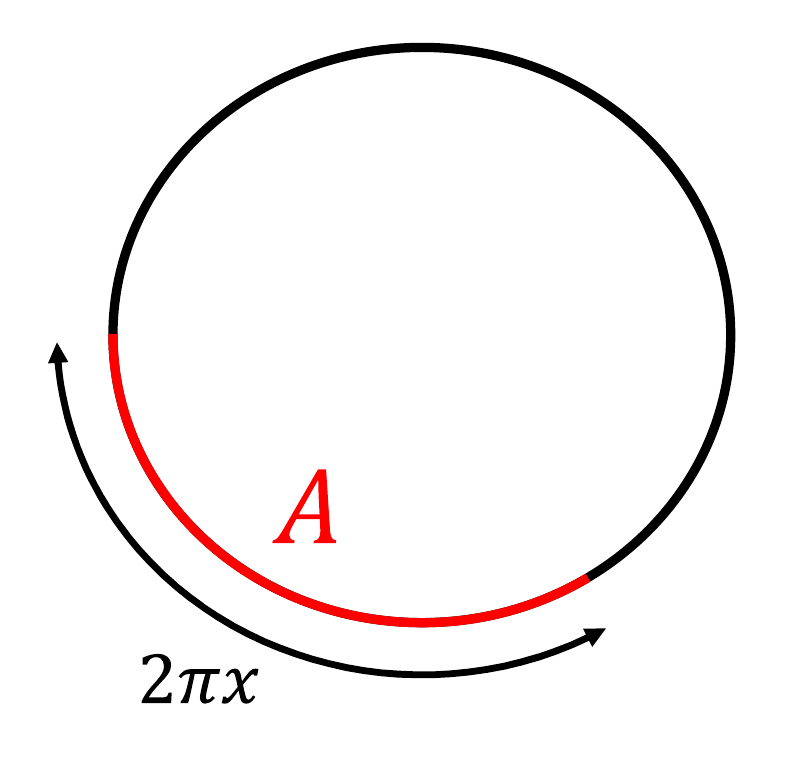}
 \end{center}
 \caption{The configuration considered for fidelity between primary states. The circumference of the system is $2\pi$ and we consider the reduced state on $A = [0,2\pi x]$.}
 \label{fig:circle}
\end{figure}

We conclude with a further application of the techniques we developed for computing the fidelity in general CFTs by evaluating the fidelity between distinct primary states in CFT. In the case where the two primary operators, $\phi_p$ and $\phi_q$, have similar and large dimensions, this probes eigenstate thermalization. If a given system exhibits eigenstate thermalization, the matrix elements of (few-body) observables, $\mathcal{O}$, in the energy eigenstate basis behave as \cite{2018RPPh...81h2001D}
\begin{equation}\label{eq:ETH}
\bra{p}\mathcal{O}\ket{q} = f_\mathcal{O}(E)\delta_{pq} + \ex{-\frac{S(E)}{2}}g_\mathcal{O}(E_p, E_q) R_{pq},
\end{equation}
where $S(E)$ is the microcanonical entropy at  $E=\frac{E_p+E_q}{2}$ and the functions $f_\mathcal{O}(E)$ and $g_\mathcal{O}(E_p,E_q)$ are smooth and ${O}(1)$.
The matrix $R_{pq}$ is a pseudo-random variable with zero mean and unit variance. We recall that if the fidelity between two state in a subregion is close to one, the trace distance must be small \cite{fuchs1999cryptographic} and then so are properly normalized matrix elements \cite{2016arXiv161000302L, 2018PhRvE..97a2140D}. This ``subsystem eigenstate thermalization'' thus implies \eqref{eq:ETH}.
 
We consider a primary state on a circle of circumference $2\pi$ reduced to a subsystem $A$ of size $2\pi x$ (see Figure \ref{fig:circle})
\begin{align}
\rho^{(p)}_A := \tr_{\bar{A}} \ket{p}\bra{p}, \quad \rho^{(q)}_A = \tr_{\bar{A}} \ket{q}\bra{q}.
\end{align}
In the short interval $(x \ll 1)$ limit, the calculation is essentially the same as that in Section \ref{uhl_subsec} such that
\begin{align}
\label{general_prim_fide}
F(\rho^{(p)}_A,\rho^{(q)}_A) = 1-
\pa{
C_{ppr}-C_{qqr}
}^2
\frac{ \Gamma\pa{\frac{1}{2} + \Delta_r } }{8 \sqrt{\pi} \Gamma\pa{1 + \Delta_r }  } (\sin \pi x)^{2\Delta_r},
\end{align}
where, as before, the subscript $r$ represents the lightest primary field in the OPE of $\phi_p \times \phi_p$ and $\phi_q \times \phi_q$. This answer holds for any CFT. We compare now with the one example in the literature that has been explicitly computed, the $c = 1$ free boson \cite{Lashkari2014}.
The fidelity between the vacuum state and the vertex state corresponding to the operator $V=\ex{i \alpha \phi}$, where $\phi$ is the boson field, is given by \cite{Lashkari2014}
\begin{equation}
F(\rho^{(0)}_A,\rho^{(V)}_A) = \cos \pa{\frac{\pi x}{2}}^{\frac{\alpha^2}{2}}. 
\end{equation}
In the short interval limit, this is approximated by
\begin{equation}\label{eq:free}
F(\rho^{(0)}_A,\rho^{(V)}_A) \simeq 1 -  \left(\frac{\pi \alpha x}{4}\right)^2.
\end{equation}
The first non-vacuum state is given by the $U(1)$ current $i\partial \phi(z)$ of conformal weight $(1,0)$,
so using the fact that $C_{VV i\partial \phi(z)} = -\alpha$, we find complete consistency with our general answer (\ref{general_prim_fide}). 

Of course, there is no exponential suppression in this fidelity, which may be understood both because the free boson theory is integrable and that here we have only considered low-lying states. In contrast, for high-energy states and irrational (such as holographic) CFTs, the OPE coefficients may be expected to lead to exponential suppression \cite{2017JHEP...05..160K,2018arXiv180409658H, 2022arXiv220306511C}.

\paragraph{Acknowledgements}
We thank Nathan Benjamin, Kanato Goto, and Juan Maldacena for useful discussions. YK is supported by the Brinson Prize Fellowship at Caltech and the U.S. Department of Energy, Office of Science, Office of High Energy Physics, under Award Number DE-SC0011632.
JKF is supported by the Institute for Advanced Study and the National Science Foundation under Grant No. PHY-2207584.

\appendix

\section{Torus four-point function}
\label{detail_GFF}

To evaluate a correlation function \eqref{sigma22_eq} on the replica manifold $\Sigma_{2,2}$, it is useful to find a conformal transformation from a torus $\bb{T}^2$ to $\Sigma_{2,2}$.
We focus on a particular map from $\bb{T}^2$ with generators of the lattice, $(2\omega_1, 2\omega_3)$ to $\Sigma_{2,2}$ with two branch cuts $A=[0,x]\cup[1,\infty]$.
This conformal map can be expressed in terms of the Weierstrass elliptic function $\wp$ as (see e.g.~\cite{olver2010nist})
\begin{align}
z = f(t) \equiv \frac{\wp(t)-e_3}{e_1-e_3},
\end{align}
where $z,t$ describe the coordinate of $\Sigma_{2,2}$ and $\bb{T}^2$.
Here the constants $e_i$ (called the lattice roots) are given by
\begin{align}\label{eq:roots}
e_i = \wp(\omega_i) \quad (i=1,2,3),
\end{align}
where $\omega_1+\omega_2+\omega_3=0$. 
The lattice roots can be expressed in terms of the Jacobi theta function $\theta_i$ or the elliptic integral of the first kind $K$ as
\begin{align}
\begin{aligned}
e_1&=\frac{\pi^2}{12\omega_1^2} \pa{\theta_2(\tau)^4 + 2 \theta_4(\tau)^4 }  = \frac{K(x)^2}{3\omega_1^2}(2-x), \\
e_2&=\frac{\pi^2}{12\omega_1^2} \pa{\theta_2(\tau)^4 -  \theta_4(\tau)^4 }  = \frac{K(x)^2}{3\omega_1^2}(2x-1), \\
e_3&=-\frac{\pi^2}{12\omega_1^2} \pa{2\theta_2(\tau)^4 + \theta_4(\tau)^4 }  = -\frac{K(x)^2}{3\omega_1^2}(1+x),
\end{aligned}
\end{align}
where the moduli parameter is defined by $\tau \equiv \frac{\omega_3}{\omega_1}$.
Here we take the following convention,
\begin{align}
K(x) = \int^{\frac{\pi}{2}}_0 \frac{d\theta}{\sqrt{1-x\sin^2\theta}} = {}_2F_1\pa{\frac{1}{2},\frac{1}{2},1;x}.
\end{align}
For this convention, the relation between the cross ratio and the moduli parameter is
\begin{align}
x=\frac{e_2-e_3}{e_1-e_3}=\pa{\frac{\theta_2(\tau)}{\theta_3(\tau)}}^4, \quad 
\tau = i \frac{K(1-x)}{K(x)}.
\end{align}
One can see using (\ref{eq:roots}) that the edges of the intervals in $\Sigma_{2,2}$ come from the generators,
\begin{align}
(\omega_1, \omega_2, \omega_3, 0)_{\bb{T}^2} \mapsto (1,x,0,\infty)_{\Sigma_{2,2}}.
\end{align}
The Weierstrass elliptic function is related to the Jacobi elliptic function as
\begin{align}
\wp(t+\omega_3)-e_3 = \pa{\frac{K(x)}{\omega_1} \sn\pa{K(x)t/\omega_1,\sqrt{x} }    }^2 x.
\end{align}
Using this relation, we can give the conformal map $g(z,x)$ from $\Sigma_{2,2}$ to $\bb{T}^2$ as
\begin{align}
t=\omega_3 + \frac{\omega_1}{K(x)} \sn^{-1} \pa{\sqrt{\frac{z}{x}},\sqrt{x}}
=\omega_3 + \frac{\omega_1}{K(x)} \int^{\sqrt{\frac{z}{x}}}_0 \frac{d\eta}{\sqrt{(1-\eta^2)(1-x\eta^2)}}
(\equiv \omega_3 + g(z,x)).
\end{align}

Now we can evaluate the geometric-mean fidelity as a correlation function on a torus.
We assume that the CFT is holographic and the operators $\phi_p, \phi_q$ are light enough to not backreact on the gravity i.e.~$1\ll h_p, h_q \ll c$.
More precisely, we assume that the conformal dimension of $\phi_p, \phi_q$ behaves like $\epsilon c$ with $\epsilon \ll 1$ in the large $c$ limit.
Then, we can evaluate the correlation function by using Wick's theorem. As a result, we obtain
that before Page time (corresponding the Hawking-Page transition of the torus partition function)
\begin{align}
\begin{aligned}
&I(\rho^p_A,\rho^q_A) \simeq  \abs{\sin \pa{\pi \pa{g(y,x) - g(y^*,x) } }}^{2h_p+2h_q}  \\
&\times
\left( 
\abs{\sin \pa{\pi\pa{g(y,x) - g(y^*,x) } }}^{2h_p+2h_q} 
+\abs{\sin \pa{\pi\pa{g(y,x) + g(y^*,x) }} }^{2h_p+2h_q}
\right)^{-1},
\end{aligned}
\end{align}
and after Page time \begin{equation}
\begin{aligned}
&I(\rho^p_A,\rho^q_A) \simeq  \abs{\sin \pa{\frac{\pi}{2\tau} \pa{g(y,x) - g(y^*,x) } }}^{2h_p+2h_q}  \\
&\times
\left( 
\abs{\sin \pa{\frac{\pi}{2\tau}   \pa{g(y,x) - g(y^*,x) } }}^{2h_p+2h_q} 
+\abs{\sin \pa{\frac{\pi}{2\tau}   \pa{g(y,x) + g(y^*,x) }} }^{2h_p+2h_q}
\right)^{-1},
\end{aligned}
\end{equation}
where we have set $2\omega_1=1$. 

The super-fidelity is related to the GM fidelity as
\begin{align}
\begin{aligned}
F_S(\rho^p_A,\rho^q_A) =
1-\frac{{1- I(\rho^p_A,\rho^q_A)}}{\Tr \rho_A^2},
\end{aligned}
\end{align}
under the condition that $\Tr \rho_A^2 := \Tr (\rho_A^q)^2=\Tr (\rho_A^p)^2$.

\addcontentsline{toc}{section}{References}
\bibliographystyle{JHEP}
\bibliography{main}

%~~~~~~~~~~~~~~~~~~~~~~~~~~~~~~~~~~~~~~~~~~~~~~~~~~~~~~~~~~~~~~~~~~~~~

\end{document}